\documentclass[pra,final]{revtex4}%
\usepackage{amsmath}
\usepackage{amsfonts}
\usepackage{amssymb}
\usepackage{graphicx}%
\providecommand{\U}[1]{\protect\rule{.1in}{.1in}}

\begin{document}
\title{Quantum/Classical Interface: Fermion Spin}
\author{W. E. Baylis, R. Cabrera, and D. Keselica}
\affiliation{Department of Physics, University of Windsor}

\begin{abstract}
Although intrinsic spin is usually viewed as a purely quantum property with no
classical analog, we present evidence here that fermion spin has a classical
origin rooted in the geometry of three-dimensional physical space. Our
approach to the quantum/classical interface is based on a formulation of
relativistic classical mechanics that uses spinors. Spinors and projectors
arise naturally in the Clifford's geometric algebra of physical space and not
only provide powerful tools for solving problems in classical electrodynamics,
but also reproduce a number of quantum results. In particular, many properites
of elementary fermions, as spin-1/2 particles, are obtained classically and
relate spin, the associated $g$-factor, its coupling to an external magnetic
field, and its magnetic moment to Zitterbewegung and de Broglie waves. The
relationship is further strengthened by the fact that physical space and its
geometric algebra can be derived from fermion annihilation and creation
operators. The approach resolves Pauli's argument against treating time as an
operator by recognizing phase factors as projected rotation operators.

\end{abstract}

\pacs{03.65.Ta, 03.65.Fd, 03.30.+p, 03.50.De}
\maketitle

\section{Introduction}

The intrinsic spin of elementary fermions like the electron is traditionally
viewed as an essentially quantum property with no classical
counterpart.\cite{LL77,Tom97} Its two-valued nature, the fact that any
measurement of the spin in an arbitrary direction gives a statistical
distribution of either \textquotedblleft spin up\textquotedblright\ or
\textquotedblleft spin down\textquotedblright\ and nothing in between,
together with the commutation relation between orthogonal components, is
responsible for many of its \textquotedblleft nonclassical\textquotedblright%
\ properties. Nevertheless, a classical spinorial approach to the relativistic
dynamics of charged particles displays these and many other quantum-like
properties and suggests that intrinsic spin itself arises from geometric
properties of three-dimensional physical space. In particular, the two-state
property, the change in sign of the spinor under a $2\pi$ rotation, the
$g$-factor of 2, and equations for spin distributions all arise classically.
The spin itself is represented by a classical intrinsic rotation that arises
as a rotational gauge freedom in the spinorial form of the Lorentz-force
equation. The spin rotation is referred to as \textquotedblleft
intrinsic\textquotedblright\ because, in this approach, a free charge is not
modeled by a classical structure, but rather it is defined as elementary only
if it has \emph{no} discernible structure other than a rest-frame direction;
that direction is interpreted as the spin axis of the particle, but only the
particle rotates, not its distribution. There is no known classical method of
obtaining the rotation rate, but the rotation does give de Broglie waves, and
measurements of the de Broglie wavelength imply rotation rates at the
Zitterbewegung frequency. The main purpose of this paper is to present
evidence of the classical origin of spin-1/2 fermions and their relation to
the geometry of physical space that arises from the classical spinor approach
to the quantum/classical (Q/C) interface .

Studies of the Q/C interface have long been of interest, both for shedding
light on quantum processes, and also for providing insight into the
unification of quantum theory with classical relativity. Spin-1/2 systems are
basic qubits, and recent work in quantum computation and
communication\cite{NielsenChuang2000}, together with the emerging field of
spintronics\cite{Awsch2002}, has focused attention on the dynamical control of
both individual spins and spin currents. These fields stand to benefit from an
improved understanding of spin at the Q/C interface. Traditional studies of
the interface have largely concentrated on quantum systems in states of large
quantum numbers and their relation to classical chaos,\cite{Bal96} to quantum
states in decohering interactions with the environment,\cite{Zur03} or in
continuum states, where semiclassical approximations are useful.\cite{Delos72}
Our approach\cite{Bay99,Bay99a,Bay03,Bay04} is fundamentally different. We
start with a description of classical relativistic dynamics using Clifford's
geometric algebra of physical space (APS). An important tool in the algebra is
the amplitude of the Lorentz transformation that boosts the system from its
\emph{rest frame }to the lab. (In this paper by \textquotedblleft rest
frame\textquotedblright, we mean the inertial frame that is instantaneously
commoving with the particle.) This amplitude enters as a spinor in a classical
context, one that satisfies linear equations of evolution admitting
superposition and interference. We explore its close relation to the quantum
wave function.

Although APS is the Clifford algebra generated by a three-dimensional
Euclidean space, it contains a four-dimensional linear space with a Minkowski
spacetime metric that allows a covariant description of relativistic
phenomena. The relativistic, spinorial treatment of APS is crucial in relating
the classical and quantum formalisms. A number of fermionic, spin-1/2,
properties follow from the spinor description of spatial rotations, and the
extension to Lorentz transformations yields immediately the momentum-space
form of the Dirac equation.\cite{Bay92a} The formulation can be extended to
the differential form of the Dirac equation and the Schr\"{o}dinger equation
in its low-velocity limit by considering a superposition of de Broglie waves.
Because of its relativistic formulation, the classical spinor description of
APS promises useful insights to the Q/C interface that may be useful not only
in quantum information theory, quantum computation, and spintronics, but also
in the foundations of quantum theory and some formulations of quantum gravity.

The association of spin-1/2 systems, and more generally of two-level systems,
to three-dimensional space is not new. States of a spin-1/2 system are
commonly represented by points on the surface of the Bloch sphere if they are
pure and inside the sphere if they are
mixed.\cite{NielsenChuang2000,Eberly2006,Fano57} Analogously, light
polarization is represented by points on or inside the Poincar\'{e}
sphere.\cite{Bay93} The three-dimensional space of these spheres is often
viewed as abstract with only indirect connections to physical space, but in
spin-1/2 systems, the direction of the Bloch vector is the same as the
polarization vector in physical space. It is also well known\cite{Corn84} that
the group $SU(2)$ of unimodular, unitary, $2\times2$ matrices, whose
representation is carried by Pauli spinors, is the universal two-fold cover of
the rotation group $SO(3)$ in three dimensions and is isomorphic to $Spin(3)$.
\cite{Loun2001} We argue here that these associations are more than
mathematical coincidence. The evidence suggests that fermionic spin-1/2
properties are inherent in the geometry of three-dimensional space. The paper
resurrects the old idea of Kronig, Uhlenbeck, and Goudsmit\cite{Tom97} that
electron represents a spinning charge, but now in a geometrical approach that
avoids the problems with superluminal velocities inherent in naive physical
models and that automatically includes Thomas-precession effects.\cite{Bay99}

Much of the evidence presented here in APS can also be formulated other
geometric algebras. Indeed, related work has been reported formulating Dirac
theory in complex quaternions\cite{Gsp01} and the spacetime algebra (STA) of
Hestenes\cite{Hes66,Hes73,Hes75,Hes79,Hes90,Hes03} and coworkers.\cite{Dor03}
While each has its particular advantages and drawbacks,\cite{BaySob03} they
share a common (isomorphic) spinorial basis for a geometric algebra describing
Lorentz transformations in spacetime. Of the three, APS is most intimately
tied to the spatial vectors and their associated geometry. It is half the size
of STA on the one hand, but unlike complex quaternions, is readily extended by
complexification. Hestenes and Gurtler\cite{HesGur71} have formulated the
Pauli theory of the electron in APS and studied spin and other local variables
in the nonrelativistic limit, in which their spinor is a scaled spatial
rotation with no boost factor. They found that Planck's constant enters the
theory only in connection with the magnitude of the spin and that the unit
imaginary $i$ in the theory generally represents the spin plane.

This paper thus collects and extends evidence that fermionic spin-1/2
properties arise from the geometry of physical space. It is important to pull
together various strands of evidence in order to minimize the danger of
ascribing physical significance to what might be mere mathematical
coincidences. While some of this evidence has been reported previously, we
believe that many of the results presented here are new, in particular (1) a
classical calculation of the magnetic moment of an electron or other
elementary fermion and its relation to Zitterbewegung, which results from (2)
a new approach to the magnetic interaction of the spin and an external
magnetic field, a calculation that supports the interpretation of the quantum
phase as an intrinsic rotation angle, that connects such rotation to de
Broglie waves, and that reveals mass as the source of energy for a spin
accelerating in a magnetic-field gradient, (3) the use of the magnetic-moment
calculation to resolve the old issue of a time operator and especially Pauli's
argument\cite{Pauli58} against such, (4) the derivation of quantum formulas
for spin distributions from classical spinors, (5) the introduction of a
conserved covariant Pauli-Luba\'{n}ski spin current, and (6) the generation of
APS as well as other Clifford algebras from fermion annihilation and creation operators.

In the next section, we see how APS arises as the natural algebra of vectors
in three-dimensional physical space, but how it also includes a
four-dimensional linear space that models spacetime of special relativity. In
Section III, the eigenspinor is introduced and shown to be a valuable tool for
finding the relativistic dynamics of particles. Although we consider it a
classical object, it is an amplitude for particle dynamics and obeys linear
equations of evolution as in quantum mechanics. Its transformation properties
show that it changes sign under a rotation of $2\pi.$ Section IV introduces
the concept of an \textquotedblleft elementary\textquotedblright\ particle in
a classical context and shows free elementary particles must be structureless
although they generally have an orientation, and if charged, they have
$g$-factors of $2$. Spin as a rotational gauge freedom is also discussed
together with its relation to de Brogile waves and Zitterbewegung. The
classical Dirac equation is derived and shown to be identical to the
momentum-space Dirac equation of quantum theory, including representations of
Dirac's gamma matrices. In Section V, classical spin distributions are shown
to be identical to quantum expressions. A simple mathematical description of
the Stern-Gerlach experiment in terms of projectors as filters is given to
show how the measurement process for the spin gives only spin-up and spin-down
states, and how the uncertainty relation for spin components arise even if the
spin itself has a well-defined direction. In the concluding section,
consequences of the eigenspinor approach to spin are summarized and several
possible extensions are briefly mentioned, including comparisons of classical
solutions of the Lorentz-force for a charge with spin to quantum solutions of
the Dirac equation, and the treatment of multiple-particle systems.

\section{The Algebra of Physical Space (APS)}

One of the simplest ways to motivate the use of Clifford's geometric algebra
is to think of vectors as square matrices. It is more common, of course, to
represent vectors as column matrices. In 3-dimensional Euclidean space, the
fixed orthonormal basis vectors are commonly written,%
\begin{equation}
\mathbf{e}_{1}=\left(
\begin{array}
[c]{c}%
1\\
0\\
0
\end{array}
\right)  ,~\mathbf{e}_{2}=\left(
\begin{array}
[c]{c}%
0\\
1\\
0
\end{array}
\right)  ,~\mathbf{e}_{3}=\left(
\begin{array}
[c]{c}%
0\\
0\\
1
\end{array}
\right)  ,
\end{equation}
and the general vector is a linear combination of these. However, the
vector-space properties are equally well served by a representation of vectors
as square matrices, using for example the Pauli spin matrices
\begin{equation}
\mathbf{e}_{1}=\left(
\begin{array}
[c]{cc}%
0 & 1\\
1 & 0
\end{array}
\right)  ,~\mathbf{e}_{2}=\left(
\begin{array}
[c]{cc}%
0 & -i\\
i & 0
\end{array}
\right)  ,~\mathbf{e}_{3}=\left(
\begin{array}
[c]{cc}%
1 & 0\\
0 & -1
\end{array}
\right)  . \label{std_rep}%
\end{equation}
In a square-matrix representation, vectors can not only be added and scaled,
they can also be multiplied together in an associative product. The vectors
together with all their products form a \emph{vector algebra}%
.\cite{Bay80,BayJones88} It becomes Clifford's \emph{geometric algebra of
physical space} (or equivalently, generated by the orthonormal basis $\left\{
\mathbf{e}_{1},\mathbf{e}_{2},\mathbf{e}_{3}\right\}  $)\emph{ }if we
constrain the product by one axiom, namely that the product of a vector with
itself is its square length:%
\begin{equation}
\mathbf{vv}\equiv\mathbf{v}^{2}=\mathbf{v\cdot v.} \label{axiom}%
\end{equation}
In particular, $\mathbf{e}_{j}^{2}=1.$ Putting $\mathbf{v=u+w,}$ we see that
$\mathbf{uw+wu}=2\mathbf{u\cdot w.}$ Thus, the familiar dot product of any two
vectors is the symmetric part of the algebraic product. Furthermore,
perpendicular vectors anticommute.

This approach also works to generate geometric algebras for pseudo-Euclidean
spaces of higher dimensions, although in some cases, such as 5-dimensional
Euclidean space, we need the additional assumption that the algebra is not
equivalent to one generated by a smaller number a basis
vectors.\cite{Loun2001} It should be emphasized that there are many possible
matrix representations for the algebra in addition to the standard one
(\ref{std_rep}). What they all share is their algebra, which is determined by
the axiom (\ref{axiom}) together with the fact that the vectors and their
products add and multiply the way square matrices of any given size do.

The product of a pair of perpendicular vectors is called a \emph{bivector }and
represents the plane containing the vectors\emph{.} For example,
$\mathbf{e}_{1}\mathbf{e}_{2}=-\mathbf{e}_{2}\mathbf{e}_{1}$ is a bivector
representing the plane containing $\mathbf{e}_{1}$ and $\mathbf{e}_{2}.$ More
importantly, it is an operator that rotates vectors in the plane by $\pi/2$.
Thus, if $\mathbf{v}=v_{x}\mathbf{e}_{1}+v_{y}\mathbf{e}_{2},$ then
$\mathbf{ve}_{1}\mathbf{e}_{2}=v_{x}\mathbf{e}_{2}-v_{y}\mathbf{e}_{1}$ is
$\mathbf{v}$ rotated counterclockwise in the plane by a right angle. To rotate
in the plane by an angle $\theta,$ we use $\mathbf{v}\exp\left(
\mathbf{e}_{1}\mathbf{e}_{2}\theta\right)  =\mathbf{v}\left(  \cos
\theta+\mathbf{e}_{1}\mathbf{e}_{2}\sin\theta\right)  ,$ whose Euler-like
expansion follows from the relation $\left(  \mathbf{e}_{1}\mathbf{e}%
_{2}\right)  ^{2}=-1.$ We say that $\mathbf{e}_{1}\mathbf{e}_{2}$ is a unit
bivector that generates rotations in the $\mathbf{e}_{1}\mathbf{e}_{2}$ plane.

\subsection{Rotations and Duals}

More general rotations of a vector $\mathbf{v}$ that may have components
perpendicular to the plane of rotation are realized by
\begin{equation}
\mathbf{v}\rightarrow R\mathbf{v}R^{\dag}, \label{rot_vector}%
\end{equation}
where the \emph{rotor} $R=\exp\boldsymbol{\Theta}=\cos\Theta+\boldsymbol{\hat
{\Theta}}\sin\Theta$ is the exponential of a bivector $\mathbf{\Theta}$ that
gives the plane of rotation and whose magnitude $\Theta$ equals the area swept
out by a unit vector in the plane by the rotation. That area is twice the
rotation angle. The dagger ($\dag$) on the second $R$ indicates a conjugation
called \emph{reversion}: it reverses the order of all vectors in the term For
example, given any two vectors $\mathbf{u,w,}$ $\left(  \mathbf{uw}\right)
^{\dag}=$ $\mathbf{wu.}$ More generally, reversion is equivalent to Hermitian
conjugation in standard representations of APS, where the basis vectors
$\mathbf{e}_{k}$ are represented by Hermitian matrices. Bivectors change sign
under reversion,and it follows that rotors are unitary: $R^{\dag}=R^{-1}.$

The product $\mathbf{e}_{1}\mathbf{e}_{2}\mathbf{e}_{3}$ is the volume element
of physical space and is called the \emph{pseudoscalar} of APS. It commutes
with all elements and squares to $\left(  \mathbf{e}_{1}\mathbf{e}%
_{2}\mathbf{e}_{3}\right)  ^{2}=-1.$ It is a \emph{trivector}, a multivector
of grade 3, whereas bivectors have grade 2, vectors grade 1, and scalars grade
0. We can identify $\mathbf{e}_{1}\mathbf{e}_{2}\mathbf{e}_{3}$ as the unit
imaginary: $\mathbf{e}_{1}\mathbf{e}_{2}\mathbf{e}_{3}=i~.$ Bivectors are then
imaginary vectors (pseudovectors) normal to the plane, for example
$\mathbf{e}_{1}\mathbf{e}_{2}=\mathbf{e}_{1}\mathbf{e}_{2}\mathbf{e}%
_{3}\mathbf{e}_{3}=i\mathbf{e}_{3}$~.This establishes a \emph{duality} between
vectors and bivectors in APS, which we can use to express rotors in terms of
the axes of rotation (but only in three dimensions): $R=\exp\left(
\mathbf{e}_{2}\mathbf{e}_{1}\phi/2\right)  =\exp\left(  -i\mathbf{e}_{3}%
\phi/2\right)  .$ The even elements (grades 0 and 2) of APS form a subalgebra
isomorphic to the quaternions.

\subsection{Spinors, Projectors, and Spin-1/2}

Rotors give a \emph{spinor} representation of rotations in a classical
context. To see what this implies, note that to combine the effect of several
rotations, we simply multiply the rotors. Thus, $R=R_{2}R_{1}$ is the rotor
for a rotation given by $R_{1}$ followed by one given by $R_{2}.$ The rotors
are elements of a group called $Spin\left(  3\right)  ,$ which is isomorphic
to $SU\left(  2\right)  $ and the universal double covering group of the
orthogonal rotation group $SO\left(  3\right)  $: $R\in Spin\left(  3\right)
\simeq SU\left(  2\right)  \simeq SO\left(  3\right)  \times Z_{2}%
~.$\cite{Loun2001}

The orientation of a system with respect to a reference frame is given by a
spinor, say the rotor $R_{1}.$ A further rotation can be viewed as a
transformation of the spinor:%
\[
R_{1}\rightarrow R=R_{2}R_{1}.
\]
This is a spinorial transformation, which has a simpler form than the general
one (\ref{rot_vector}) for vectors. Note that the rotation of a spinor by
$2\pi$ about any axis $\mathbf{m}$ introduces a factor of $\exp\left(
-i\pi\mathbf{m}\right)  =-1.:$This is also a distinguishing property of
fermions. The rotor $R$ is a \emph{reducible }rotational spinor.
\emph{Irreducible spinors }are formed by applying $R$ to a \emph{projector} (a
real idempotent), such as
\begin{equation}
P_{3}\equiv\frac{1}{2}\left(  1+\mathbf{e}_{3}\right)  =P_{3}^{2}=P_{3}^{\dag
}. \label{P3}%
\end{equation}
The irreducible spinor is $R$P$_{3},$ and twice its even-grade part is $R.$
Note, however, that projectors and irreducible spinors are not invertible. The
existence of noninvertible elements demonstrates that APS is not a division
algebra. Instead, it embodies a rich mathematical framework that admits such
proven powerful tools as projectors and spinors, even in classical physics.

Elements that can be written in the form $x$P$_{3},$ where $x$ is an arbitrary
element of APS, are said to lie within a \emph{minimal left ideal }of APS that
we denote $\left(  \text{APS}\right)  $P$_{3}.$ The elements of a minimal left
ideal can all be expressed as even elements of the algebra times the projector
of the ideal. Even elements of APS are quaternions, which may be considered
spatial rotors times scalar dilation factors, and they do form a division
algebra. The proof that any element of $\left(  \text{APS}\right)  $P$_{3}$ is
equivalent to an even element of APS times P$_{3}$ is a trivial result of the
\textquotedblleft pacwoman\textquotedblright\ property\cite{Bay99} that
$\mathbf{e}_{3}$P$_{3}=$P$_{3}.$ Elements of $\left(  \text{APS}\right)
$P$_{3}$ can be specified by two complex-valued functions and are thus
appropriate for a Hilbert-space treatment of quantum mechanics. In particular,
the rotor for a rotation expressed in Euler angles can be written%
\begin{equation}
R=\exp\left(  \mathbf{e}_{21}\phi/2\right)  \exp\left(  \mathbf{e}_{13}%
\theta/2\right)  \exp\left(  \mathbf{e}_{21}\chi/2\right)  , \label{Euler1}%
\end{equation}
and its projected ideal form is%
\begin{equation}
R\text{P}_{3}=\left(  e^{-i\phi/2}\cos\frac{\theta}{2}+\mathbf{e}_{1}%
e^{i\phi/2}\sin\frac{\theta}{2}\right)  e^{-i\chi/2}\text{P}_{3}~.
\end{equation}
The standard matrix representation is the usual two-component rotational
spinor traditionally associated with a spin-1/2 system plus a second column of
zeros:%
\begin{equation}
R\text{P}_{3}=e^{-i\chi/2}\left(
\begin{array}
[c]{cc}%
e^{-i\phi/2}\cos\theta/2 & 0\\
e^{i\phi/2}\sin\theta/2 & 0
\end{array}
\right)  . \label{RP_3}%
\end{equation}
All elements of the minimal left ideal $\left(  \text{APS}\right)  $P$_{3}$
have a vanishing second column in the standard representation.

\subsection{Physical Space from Fermion Annihilation/Creation Operators}

Another way to relate physical space to fermions is to generate APS from a
pair $a,a^{\dag}$ of annihilation and creation operators for a given state.
These are defined to be nilpotent elements $a^{2}=0=\left(  a^{\dag}\right)
^{2}$ that satisfy the anticommutation relation%
\begin{equation}
aa^{\dag}+a^{\dag}a=1. \label{unity}%
\end{equation}
In addition to (\ref{unity}), the only other real elements that can be
generated from $a,a^{\dag}$ are%
\begin{align}
\mathbf{e}_{1}  &  =a+a^{\dag}\label{eac_relations}\\
\mathbf{e}_{2}  &  =i\left(  a-a^{\dag}\right) \nonumber\\
\mathbf{e}_{3}  &  =a^{\dag}a-aa^{\dag}\nonumber
\end{align}
and these satisfy the Clifford relations $\mathbf{e}_{j}\mathbf{e}%
_{k}+\mathbf{e}_{k}\mathbf{e}_{j}=2\delta_{jk}$ from which APS is derived. The
relations (\ref{eac_relations}) can be solved for $a,a^{\dag}$ to give%
\begin{align}
a  &  =\frac{1}{2}\left(  \mathbf{e}_{1}-i\mathbf{e}_{2}\right)
=\mathbf{e}_{1}\text{P}_{3}\nonumber\\
a^{\dag}  &  =\text{P}_{3}\mathbf{e}_{1}=\mathbf{e}_{1}\text{\={P}}_{3}%
\end{align}
which identifies the annihilation and creation operators as null flags of
APS.\cite{Bay99,PenroseRindler} Taken together with P$_{3}=aa^{\dag}$ and
\={P}$_{3}=a^{\dag}a,$ they form a null basis for spacetime. Note that one may
also view $a^{\dag},a$ as raising and lowering operators for a spin-1/2 system.

Fermion annihilation/creation operators for two or more states can be used to
generate geometric algebras for high-dimensional spaces. For example, the
orthonormal vectors of 4-dimensional Euclidean space can be expressed in terms
of annihilation/creation operators for two fermion states:
\begin{align}
\mathbf{e}_{1}  &  =a_{1}^{\dagger}+a_{1}\\
\mathbf{e}_{2}  &  =i(a_{1}-a_{1}^{\dagger})\\
\mathbf{e}_{3}  &  =a_{2}^{\dagger}+a_{2}\\
\mathbf{e}_{4}  &  =i(a_{2}-a_{2}^{\dagger})
\end{align}
The full geometric algebra can be generated from these basis vectors. Rotors
in the four-dimensional space are elements of $Spin(4)\simeq SU\left(
2\right)  \otimes SU\left(  2\right)  $ and can be described by two
independent spin-1/2 systems or qubits. However, rotors in four dimensions,
generated by bivectors, do not span the whole Hilbert space, as shown in
\cite{Cabrera07}. In higher dimensions more fermion states are required.

\section{Paravectors and Spacetime}

Every element of APS is some real linear combination of scalars (grade 0),
vectors (grade 1), bivectors (grade 2), and trivectors (grade 3). However, as
seen above, trivectors are pseudoscalars, which are expressed as imaginary
scalars, and bivectors are pseudovectors, expressed as imaginary vectors.
Thus, every element of APS is a linear combination of a complex scalar and a
complex vector. The algebra of real vectors in three dimensions thereby forms
a complex linear space of four dimensions. This possibility was actually
evident from the $2\times2$ matrix representation of our original vectors. The
vector elements of the four-dimensional space are called \emph{paravectors}%
\cite{Bay99,Loun2001} to distinguish them from vectors of the original real
three-dimensional space. A paravector is the sum of a scalar and a vector, for
example $p=p^{0}+\mathbf{p,}$ where $p^{0}$ is the scalar part. To reinforce
the four-dimensional property of paravector space, we write $p=p^{\mu}e_{\mu
},$ where our paravector basis elements are $e_{0}=1$ and $e_{k}%
=\mathbf{e}_{k},~k=1,2,3,$ and we adopt the Einstein summation convention of
summing over indices that appear once as an upper index and once as a lower
one. Note that APS also contains several linear subspaces of interest: the
center of the algebra comprises scalars plus pseudoscalars: the \emph{complex
numbers}; the elements of even grade (scalars plus bivectors, the even
subalgebra of APS) are \emph{quaternions}; and elements of grades 0 and 1 are
\emph{real paravectors}.

The metric of paravector space suggests the physical significance of the
fourth dimension. The original three-dimensional space has a Euclidean metric,
and the paravector metric is determined by an appropriate quadratic form, that
is a scalar expression representing the square length of a paravector. In the
original vector space, the quadratic form was identified as the square of the
vector, but the square of a paravector is not a scalar. As with complex
numbers, we must multiply the paravector by a conjugate to be sure of getting
a scalar. The appropriate conjugate is the \emph{Clifford conjugate} $\bar
{p}=p^{0}-\mathbf{p}$ because%
\begin{equation}
p\bar{p}=\bar{p}p=\left(  p^{0}\right)  ^{2}-\mathbf{p}^{2}%
\end{equation}
is always a scalar. It can be adopted as the quadratic form. As long as the
quadratic form $x\bar{x}$ of an element $x$ of APS does not vanish, the
inverse of $x$ is%
\[
x^{-1}=\bar{x}\left(  x\bar{x}\right)  ^{-1}=\left(  x\bar{x}\right)
^{-1}\bar{x}.
\]
The quadratic form $x\bar{x}$ of any element $x$ equals the determinant of its
matrix representation. For consistency, the product of any two elements $x,y,$
has the Clifford conjugate $\overline{xy}=\bar{y}\bar{x}.$

We can use the Clifford conjugate to isolate the scalar-like (S) and
vector-like (V) parts of any element $p,$ and in the same way reversion
($\dag$) can be used to separate the \textquotedblleft real\textquotedblright%
\ ($\Re,$ or hermitian) and \textquotedblleft imaginary\textquotedblright%
\ ($\Im,$ or antihermitian) parts:%
\begin{align}
\left\langle p\right\rangle _{S}  &  \equiv\frac{1}{2}\left(  p+\bar
{p}\right)  ,\ \left\langle p\right\rangle _{V}\equiv\frac{1}{2}\left(
p-\bar{p}\right) \nonumber\\
\left\langle p\right\rangle _{\Re}  &  \equiv\frac{1}{2}\left(  p+p^{\dag
}\right)  ,\ \left\langle p\right\rangle _{\Im}\equiv\frac{1}{2}\left(
p-p^{\dag}\right)  .
\end{align}
The scalar-like part of any element is half the trace of its matrix
representation, and for any two elements $p,x,$ $\left\langle px\right\rangle
_{S}=\left\langle xp\right\rangle _{S}.$One can easily verify that $p\bar{p}$
is its own Clifford conjugate. If we replace $p$ by the sum $p+q$ of two
paravectors, we can determine the scalar product of $p$ with $q$:%

\begin{equation}
\left\langle p\bar{q}\right\rangle _{S}\equiv\frac{1}{2}\left(  p\bar{q}%
+q\bar{p}\right)  =p^{\mu}q^{\nu}\left\langle e_{\mu}\bar{e}_{\nu
}\right\rangle _{S}=p^{\mu}q^{\nu}\eta_{\mu\nu}. \label{scalar}%
\end{equation}
The tensor $\left(  \eta_{\mu\nu}\right)  =\left(  \left\langle e_{\mu}\bar
{e}_{\nu}\right\rangle _{S}\right)  =$ diag$\left(  1,-1,-1,-1\right)  $ is
the metric tensor of Minkowski spacetime. Had we chosen the quadratic form to
be $-p\bar{p},$ we would still have arrived at a Minkowski spacetime metric,
but one with the opposite signature. In either case, we see that real
paravectors can represent vectors in flat four-dimensional spacetime. Two
paravectors are \emph{orthogonal} if their scalar product vanishes.

We can now extend rotors to \textquotedblleft rotations\textquotedblright\ in
spacetime. The \emph{biparavector }basis element $\left\langle e_{\mu}\bar
{e}_{\nu}\right\rangle _{V}$ generates rotations in the spacetime plane
containing the paravectors $e_{\mu}$ and $e_{\nu}.$ Lorentz rotors $L=\pm
\exp\left(  \mathbf{W}/2\right)  $ with $\mathbf{W}$ a biparavector induce
restricted Lorentz transformations of any spacetime vector $p:$
\begin{equation}
p\rightarrow LpL^{\dag}. \label{Lorentz_rot}%
\end{equation}
The rotor $L\in\$pin_{+}\left(  1,3\right)  \simeq SL\left(  2,\mathbb{C}%
\right)  \simeq SO_{+}\left(  1,3\right)  \times Z_{2}$ is an amplitude for
the Lorentz transformation.\cite{Loun2001} As with spatial rotors $R,$ the
Lorentz rotor has unit spacetime length (is \emph{unimodular}): $L\bar{L}=1$
and as with any invertible linear operator, there is a polar decomposition of
$L$ into the product of unitary and hermitian factors. Since $L$ is
unimodular, so are its factors, and we can write $L=BR,$ where $R$ is a
unitary spatial rotor and $B=B^{\dag}$ is a boost (velocity transformation).
Since the Clifford conjugate of the transformation (\ref{Lorentz_rot}) is
$\bar{p}\rightarrow\bar{L}^{\dag}\bar{p}\bar{L},$ the Lorentz transformation
of $p\bar{q},$ where $p$ and $q$ are paravectors, takes the form $p\bar
{q}\rightarrow Lp\bar{q}\bar{L},$ and it follows that the scalar product
(\ref{scalar}) is invariant.

Just as we identified multivectors of vector grades 0 to 3, we can identify
other elements as multiparavectors of paravector grades 0 to 4. The relation
is given in table \ref{grades}.
\begin{table}[tbp] \centering
\caption{Relations of paravector grade (pv-grade) to vector grade (v-grade). There exists a linear space for each vector and paravector grade. The number (no.) of independent elements is the dimension of the corresponding linear space.}
\label{grades}
\begin{tabular}
[c]{lllll}%
\textbf{pv-grade} & \textbf{pv-type} & \textbf{no.} & \textbf{v-grades} &
\textbf{basis elements}\\\hline
0 & scalar & 1 & $0$ & $1=e_{0}=\left\langle e_{0}\right\rangle _{\Re S}$\\
1 & paravector & 4 & $0+1$ & $e_{\mu}=\left\langle e_{\mu}\right\rangle _{\Re
}$\\
2 & biparavector & 6 & $1+2$ & $\left\langle e_{\mu}\bar{e}_{\nu}\right\rangle
_{V}$\\
3 & triparavector & 4 & $2+3$ & $\left\langle e_{\lambda}\bar{e}_{\mu}e_{\nu
}\right\rangle _{\Im}$ or $i\mathbf{e}_{\rho}$\\
4 & pseudoscalar & 1 & $3$ & $\left\langle e_{\lambda}\bar{e}_{\mu}e_{\nu}%
\bar{e}_{\rho}\right\rangle _{\Im S}$ or $i$%
\end{tabular}
\end{table}
Note that whereas paravector grades 1, 2, and 3 have contributions from two
neighboring vector grades (see Table \ref{grades}), the spacetime scalars are
the same as vector scalars, and that the pseudoscalar element in spacetime is
$i,$ the same as the vector pseudoscalar. This permits a simple calculation of
Hodge-type duals of elements: if $x$ is any element of APS, even a multigrade
one, its Clifford-Hodge dual\cite{Bay99,Loun2001} is defined to be $~^{\ast
}x=-ix.$

\subsection{Classical Eigenspinors for Relativistic Dynamics}

The Lorentz rotor that transforms between the particle rest frame and the lab
is useful for describing particle dynamics. With its help, the velocity and
orientation of the particle can be calculated and, indeed, any property known
in the rest frame can be transformed to the lab.\cite{Bay99,Hes99} Because of
its special status, we give this Lorentz rotor a special designation: it is
the \emph{eigenspinor }$\Lambda$ of the particle. For an accelerating
particle, $\Lambda$ is a function of the proper time $\tau$ of the particle,
representing at each instant the Lorentz rotor from the inertial frame
commoving with the particle (the \textquotedblleft rest
frame\textquotedblright) to the lab. For example, the \emph{proper velocity
}$u_{0}$ of the particle in the lab is (in units with $c=1$) just the
transformed time axis:
\begin{equation}
u_{0}=\Lambda e_{0}\Lambda^{\dag}. \label{prop_vel}%
\end{equation}
The proper velocity is a spacetime vector and can be further transformed by a
Lorentz rotor $L:$ $u_{0}\rightarrow Lu_{0}L^{\dag}=L\Lambda e_{0}%
\Lambda^{\dag}L^{\dag}.$ This is equivalent to the Lorentz rotation
$\Lambda\rightarrow L\Lambda$ of the eigenspinor. This form of a Lorentz
transformation is distinct from that (\ref{Lorentz_rot}) for a spacetime
vector or any product of vectors, and that is part of the justification for
calling $\Lambda$ a spinor.

The other basis paravectors of an elementary system can be similarly
transformed to the lab frame. The system \emph{tetrad }$\left\{  u_{\mu
}\right\}  $ is the set of transformed basis elements%
\begin{equation}
u_{\mu}=\Lambda e_{\mu}\Lambda^{\dag}.
\end{equation}
In addition to the special role played by the proper velocity $u_{0},$ the
spacetime vector $u_{3}=\Lambda e_{3}\Lambda^{\dag}$ may be identified with
the \emph{Pauli-Luba\'{n}ski (PL) spin}. Whereas $u_{0}$ is a timelike unit
paravector because $u_{0}\bar{u}_{0}=1,$ the PL spin is spacelike: $u_{3}%
\bar{u}_{3}=-1$ and is orthogonal to $u_{0}:$%
\begin{equation}
\left\langle u_{3}\bar{u}_{0}\right\rangle _{S}=\left\langle e_{3}\bar{e}%
_{0}\right\rangle _{S}=0.
\end{equation}
The spacetime dual of $u_{3}$ is%
\begin{equation}
-iu_{3}=\Lambda e_{1}\bar{e}_{2}e_{0}\Lambda^{\dag}=\mathbf{S}u_{0},
\label{PLdual}%
\end{equation}
where $\mathbf{S}\equiv\Lambda e_{1}\bar{e}_{2}\bar{\Lambda}=u_{1}\bar{u}_{2}$
is recognized as the spacetime bivector (\textquotedblleft
biparavector\textquotedblright) for the plane orthogonal to both $u_{0}$ and
$u_{3}.$ If $\Lambda$ is a pure spatial rotation, $u_{3}$ is simply the unit
spatial vector $\mathbf{s}=R\mathbf{e}_{3}R^{\dag}=i\mathbf{S}.$ The
association of $e_{3}$ and $u_{3}$ with spin will be made more definite below.

A system of several parts will generally require several eigenspinors to
describe its motion. A system is said to be \emph{elementary} if all of its
motion is described by a \emph{single} $\Lambda\left(  \tau\right)  .$ A free
elementary system (\textquotedblleft particle\textquotedblright) is
necessarily unstructured, but it may have an orientation and is not
necessarily point-like.

\subsection{Equation of Motion}

Since the eigenspinor at any instant $\tau$ is a Lorentz rotor and every rotor
$L$ has an inverse $\bar{L}$, the eigenspinor at different times is related by%
\begin{equation}
\Lambda\left(  \tau_{2}\right)  =L_{2}\bar{L}_{1}\Lambda\left(  \tau
_{1}\right)  \equiv L\left(  \tau_{2},\tau_{1}\right)  \Lambda\left(  \tau
_{1}\right)  ,
\end{equation}
where $L_{1}\equiv\Lambda\left(  \tau_{1}\right)  ,~L_{2}\equiv\Lambda\left(
\tau_{2}\right)  ,$ and we noted that by their group property, the product of
Lorentz rotors is another Lorentz rotor. The proper-time derivative of the
eigenspinor can be expressed%
\begin{equation}
\dot{\Lambda}=\left(  \dot{\Lambda}\bar{\Lambda}\right)  \Lambda=\frac{1}%
{2}\boldsymbol{\Omega}\Lambda=\frac{1}{2}\Lambda\mathbf{\Omega}_{\text{rest}}
\label{Lam_dot}%
\end{equation}
with $\boldsymbol{\Omega}=2\dot{\Lambda}\bar{\Lambda},$ and it follows from
the unimodularity of $\Lambda$ that $\mathbf{\Omega}$ is a biparavector. From
the infinitesimal time development%
\begin{equation}
\Lambda\left(  \tau+d\tau\right)  =\left(  1+\frac{1}{2}\boldsymbol{\Omega
~}d\tau\right)  \Lambda\left(  \tau\right)  =\exp\left(  \frac{1}%
{2}\boldsymbol{\Omega}~d\tau\right)  \Lambda\left(  \tau\right)  ,
\end{equation}
one can interpret $\boldsymbol{\Omega}$ as the \emph{spacetime rotation rate
}of the particle frame. Similarly, $\mathbf{\Omega}_{\text{rest}}=\bar
{\Lambda}\mathbf{\Omega}\Lambda$ is the rotation rate in the rest frame of the
particle. If $\mathbf{\Omega}$ is known, we can find the proper time-rate of
change of any property known in the rest frame. For example, the proper
acceleration of the particle is given by
\begin{equation}
\dot{u}_{0}=\dot{\Lambda}e_{0}\Lambda^{\dag}+\Lambda e_{0}\dot{\Lambda}^{\dag
}=\left\langle \boldsymbol{\Omega}u_{0}\right\rangle _{\Re}. \label{u_dot}%
\end{equation}

\subsection{Maxwell-Lorentz Theory}

Comparison of $\dot{u}_{0}$ (\ref{u_dot}) to the Lorentz-force equation
$\dot{p}=m\dot{u}_{0}=e\left\langle \mathbf{F}u_{0}\right\rangle _{\Re}$
suggests a covariant definition of the electromagnetic field as the spacetime
rotation rate per unit charge-to-mass ratio%
\begin{equation}
\mathbf{F}=\left\langle \partial\bar{A}\right\rangle _{V}=\mathbf{E+}%
i\mathbf{B}=m\boldsymbol{\Omega}/e. \label{Field}%
\end{equation}
It also gives a spinor form of Lorentz-force equation:
\begin{equation}
\dot{\Lambda}=\frac{e}{2m}\mathbf{F}\Lambda\label{Lor_force}%
\end{equation}
that simplifies many problems in electrodynamics. For example, if $\mathbf{F}$
is constant, we can integrate (\ref{Lor_force}) immediately to get the
eigenspinor $\Lambda\left(  \tau\right)  =\exp\left(  \frac{e}{2m}%
\mathbf{F}\tau\right)  \Lambda\left(  0\right)  ,$ which determines both the
proper velocity (\ref{prop_vel}) of the particle and the orientation of its
reference frame. The spinor equation (\ref{Lor_force}) also reveals surprising
symmetries, for example, the fact that the field $\mathbf{F}_{\text{rest}}$
seen in the instantaneous rest frame of the particle is constant, even if the
particle is accelerating: $\mathbf{F}_{\text{rest}}\left(  \tau\right)
=\bar{\Lambda}\left(  \tau\right)  \mathbf{F}\Lambda\left(  \tau\right)
=\mathbf{F}_{\text{rest}}\left(  0\right)  .$

To complete the formulation of Maxwell-Lorentz theory, we need Maxwell's
equations relating $\mathbf{F}$ to the charge-current density $j=eJ=e\rho
+\mathbf{j.}$ These, in SI units, are just the scalar, vector, pseudovector,
and pseudoscalar components of $\bar{\partial}\mathbf{F}=\mu\bar{j},$ where
$\mu$ is the permeability of space.\cite{Bay99} Simple expansions of these
algebraic equations in basis elements yield the traditional corresponding
tensor equations. The eigenspinor approach has proved to be a powerful tool
for finding exact solutions in classical relativistic
dynamics.\cite{Bay99,Bay99a}

Note that the spinor equation (\ref{Lor_force}) is invariant under a change in
the orientation of the rest frame:%
\begin{equation}
\Lambda\rightarrow\Lambda R, \label{reorient}%
\end{equation}
where $R$ is any fixed spatial rotor. The transformation (\ref{reorient}) may
be considered a global gauge transformation of $\Lambda.$ The invariance can
be extended to a \emph{local gauge transformation }$\Lambda\rightarrow
\Lambda_{\omega_{0}}=\Lambda\exp\left(  -i\mathbf{e}_{3}\omega_{0}\tau\right)
$ by adding a rotational gauge term to (\ref{Lor_force}) that represents a
rest-frame rotation or spin [see Eq. (\ref{PLdual})]:%
\begin{equation}
\dot{\Lambda}_{\omega_{0}}=\frac{e}{2m}\mathbf{F}\Lambda_{\omega_{0}}%
-i\omega_{0}\Lambda_{\omega_{0}}\mathbf{e}_{3}=\frac{1}{2}\left(  \frac{e}%
{m}\mathbf{F}+2\omega_{0}\mathbf{S}\right)  \Lambda_{\omega_{0}}~.
\label{Lor_f_w_spin}%
\end{equation}

\section{Dirac Equation in APS}

The spacetime momentum of a particle is given by $p=\Lambda m\Lambda^{\dag},$
but since the eigenspinor $\Lambda$ is an invertible Lorentz rotor, we can
equally well write the relation $p\bar{\Lambda}^{\dag}=m\Lambda,$which may be
called a real-linear form since real superpositions of solutions are also
solutions. Note that we have not assumed that the particle whose dynamics are
described by $\Lambda$ has a point-like distribution. It may indeed be
distributed in space with some density $\rho$ in its rest frame. The current
density $J$ in the lab is then $J=\Lambda\rho\Lambda^{\dag}\equiv\Psi
e_{0}\Psi^{\dag},$ where we have put $\Psi\equiv\rho^{1/2}\Lambda.$ Since
$\rho$ is a real scalar, equation $p\bar{\Lambda}^{\dag}=m\Lambda$ is also
satisfied by the current amplitude $\Psi:$%
\begin{equation}
p\bar{\Psi}^{\dag}=m\Psi. \label{class_Dirac}%
\end{equation}
This is the \emph{classical Dirac equation}.\cite{Bay92a}

To cast the equation in complex linear form required for a Hilbert-space
formulation, we project it into minimal left ideals of APS%
\begin{align}
p\bar{\Psi}^{\dag}\text{P}_{3}  &  =m\Psi\text{P}_{3}\\
p\bar{\Psi}^{\dag}\text{\={P}}_{3}  &  =m\Psi\text{\={P}}_{3}~.\nonumber
\end{align}
We can now flip the ideal of the second equation with bar-dagger conjugation
to get $\bar{p}\Psi$P$_{3}=m\bar{\Psi}^{\dag}$P$_{3}$ so that both equations
lie in the same minimal left ideal of APS. As noted above, all elements of the
ideal have only two independent complex components. If we stack them, using
the Pauli-matrix representation of APS, we get a four-component column matrix
identical to the Dirac spinor in the Weyl representation%
\begin{equation}
\psi^{\left(  W\right)  }=\frac{1}{\sqrt{2}}\left(
\begin{array}
[c]{c}%
\Psi\mathsf{P}_{+3}\\
\bar{\Psi}^{\dag}\mathsf{P}_{+3}%
\end{array}
\right)  , \label{DiracW}%
\end{equation}
and the equation for it is exactly the Dirac equation in momentum form,
complete with gamma matrices in the Weyl representation. The projection of the
algebraic $\Psi$ by $\mathsf{P}_{3}$ and \={P}$_{3}=$ P$_{-3}$ picks out the
upper and lower component pairs of $\psi^{\left(  W\right)  }$ and is seen to
be equivalent to multiplication of $\psi^{\left(  W\right)  }$ by the
traditional chirality projectors $\frac{1}{2}\left(  1\pm\gamma_{5}\right)  $
with $\gamma_{5}=-i\gamma_{0}\gamma_{1}\gamma_{2}\gamma_{3}$ $:$%
\begin{equation}
\frac{1}{2}\left(  1\pm\gamma_{5}\right)  \psi^{\left(  W\right)
}\Leftrightarrow\Psi\mathsf{P}_{\pm3}~. \label{chiral_split}%
\end{equation}
We might therefore refer to the minimal left ideals $C\!\ell_{3}%
\mathsf{P}_{\pm3}$ as the left and right chiral ideals of APS.

\subsection{De Broglie Waves and Spin Interaction}

The spin rotation of a free eigenspinor projects into a phase oscillation
$\exp\left(  -i\mathbf{e}_{3}\omega_{0}\tau\right)  $P$_{3}=e^{-i\omega
_{0}\tau}$P$_{3}.$ A spatial distribution with a synchronized phase
oscillation becomes a de Broglie wave in the lab frame after a boost by the
eigenspinor $\Lambda.$\cite{Bay07} The boost desynchronizes the phase
oscillations across the distribution, giving a wave of wavelength
$\lambda=2\pi/\left(  \gamma\left\vert \mathbf{v}\right\vert \omega
_{0}\right)  ,$ where $\mathbf{v}$ is the boost velocity and $\gamma=\left(
1-\mathbf{v}^{2}\right)  ^{-1/2}$ is the Lorentz dilation factor. In terms of
the Lorentz invariant $\tau=\left\langle x\bar{u}\right\rangle _{S},$ the
phase factor in lab coordinates is%
\begin{equation}
\exp\left(  -i\omega_{0}\tau\right)  =\exp\left(  -i\omega_{0}\left\langle
x\bar{u}_{0}\right\rangle _{S}\right)  =\exp\left[  -i\gamma\omega_{0}\left(
t-\mathbf{v\cdot x}\right)  \right]  ~.
\end{equation}
The wavelength has the measured de Broglie value $\lambda=2\pi\hbar/\left(
\gamma m\left\vert \mathbf{v}\right\vert \right)  $ if and only if the
oscillations occur at the Zitterbewegung frequency\cite{Zitterref} $\omega
_{0}=E_{0}/\hbar=m/\hbar.$ We note that for a compound system, the rest energy
$m$ includes internal motion and interaction. The frequency involved is very
rapid, even for an electron, for which $\omega_{0}\simeq0.776\times10^{21}%
\operatorname{s}%
^{-1}$ and it is therefore clear that it can only refer to an intrinsic
rotation, not to a rotation of the distribution as a whole. Any observation of
an elementary system with such an intrinsic spin can only see it as
essentially point-like, with no discernible physical extent. This is
consistent with the standard Born interpretation of the quantum wave function
as a probability amplitude.

We noted above that a magnetic field can be defined by the spatial rotation
rate it induces. For an elementary charge at rest, a pure magnetic field
$\mathbf{B}$ causes a shift in the total spatial rotation rate from
$\left\vert \mathbf{\Omega}\right\vert =2\omega_{0}$ to $\left\vert
2\omega_{0}\mathbf{\hat{s}-}\left(  e/m\right)  \mathbf{B}\right\vert
\simeq2\omega_{0}-\left(  e/m\right)  \mathbf{\hat{s}\cdot B}$ for fields
small compared to $2\omega_{0}m/e\simeq4.414\times10^{9}$ tesla for an
electron, and this shift in rotation frequency corresponds to a mass change
and hence an interaction energy $-\boldsymbol{\mu}\cdot\mathbf{B,}$ where
\begin{equation}
\boldsymbol{\mu}=\frac{e\hbar}{2m}\hat{s} \label{magdipmmt}%
\end{equation}
should evidently be interpreted as the magnetic moment of the fermion. The
$g$-factor, which gives $2m/e$ times the ratio of the magnetic moment to the
angular momentum,\cite{JDJ99} is given by the definition of an elementary
particle: since its motion is described by a single eigenspinor field
$\Lambda,$ its cyclotron and Larmor-precession frequencies must be equal. This
implies $g=2$.\cite{Bay92a} Taken together, the $g$-factor and the magnetic
moment imply a spin angular momentum of magnitude $\hbar/2$ in the direction
$\mathbf{\hat{s}.}$ The analysis not only derives the interaction of a spin in
a magnetic field, it also supports the classical picture of the spin as a
physical (but intrinsic) rotation at the rate $2\omega_{0}$ in a right-handed
sense about the direction $\mathbf{\hat{s}}=R\mathbf{e}_{3}R^{\dag}.$ The
calculation also reveals the mass as a source of the energy when the magnetic
moment of a spin is accelerated in an inhomogeneous but static magnetic field.

Our picture differs considerably from that of Hestenes\cite{Hes79,Hes90}, who
models the electron as a point charge moving at the speed of light on a
helical path that circles at the Zitterbewegung frequency.

\subsection{Large and Small Components}

For bound states and at low velocities, it is convenient to use $\left\langle
\Psi\right\rangle _{\pm}=\frac{1}{2}\left(  \Psi\pm\bar{\Psi}^{\dagger
}\right)  ,$ which are the even and odd parts of $\Psi.$ They are even and odd
not only in the Clifford algebra sense of containing only even-grade or only
odd-grade elements of APS, but also in the sense of being even and odd under
parity inversion. The even and odd parts of $\Psi$ correspond to the
\emph{large} and \emph{small} components of positive-energy solutions at low
velocities:%
\begin{align}
\left\langle \Psi\right\rangle _{+}  &  =\rho^{1/2}\left\langle B\right\rangle
_{+}R=\rho^{1/2}\sqrt{\frac{m+E}{2m}}R\simeq\rho^{1/2}R\label{Even_Psi}\\
\left\langle \Psi\right\rangle _{-}  &  =\rho^{1/2}\left\langle B\right\rangle
_{-}R=\frac{\mathbf{p}}{m+E}\left\langle \Psi\right\rangle _{+}~,
\label{Odd_Psi}%
\end{align}
where we noted that the scalar function $\rho^{1/2}$ and rotor $R$ are both
even elements and that $B=\left(  p/m\right)  ^{1/2}=\left(  p+m\right)
/\sqrt{2\left(  m+E\right)  }$~.\cite{Bay99} The third term on the RHS of
(\ref{Even_Psi}, \ref{Odd_Psi}) is for free particles of energy $E$, since for
these%
\begin{align}
B  &  =\sqrt{\frac{p}{m}}=\frac{p+m}{\sqrt{2m\left(  E+m\right)  }%
},\ \label{B_even_odd}\\
\left\langle B\right\rangle _{+}  &  =\sqrt{\frac{E+m}{2m}},\ \nonumber\\
\left\langle B\right\rangle _{-}  &  =\frac{\mathbf{p}}{\sqrt{2m\left(
E+m\right)  }}=\frac{\mathbf{p}}{E+m}\left\langle B\right\rangle
_{+}~.\nonumber
\end{align}
The last expression for $\left\langle \Psi\right\rangle _{+}$ on the RHS is
the low-velocity approximation. In the rest frame, the small component
$\left\langle \Psi\right\rangle _{-}$ disappears and the eigenfunction is
even. We say the particle has even \emph{intrinsic parity}.%
\index{parity!intrinsic}%
\emph{\ }Note that the spinors $\left\langle \Psi\right\rangle _{\pm}$ are
easily extracted from the corresponding ideal spinors $\left\langle
\Psi\right\rangle _{\pm}\mathsf{P}_{+3}:$%
\begin{equation}
\left\langle \Psi\right\rangle _{+}=2\left\langle \left\langle \Psi
\right\rangle _{+}\mathsf{P}_{+3}\right\rangle _{+},\ \left\langle
\Psi\right\rangle _{-}=2\left\langle \left\langle \Psi\right\rangle
_{-}\mathsf{P}_{+3}\right\rangle _{-}~.
\end{equation}
The Dirac bispinor in the Dirac-Pauli (or standard) representation is related
by
\begin{equation}
\psi^{\left(  DP\right)  }=\frac{1}{\sqrt{2}}\left(
\begin{array}
[c]{cc}%
1 & 1\\
1 & -1
\end{array}
\right)  \psi^{\left(  W\right)  }=\binom{\left\langle \Psi\right\rangle
_{+}\mathsf{P}_{+3}}{\left\langle \Psi\right\rangle _{-}\mathsf{P}_{+3}}\,,
\label{DiracDP}%
\end{equation}

Generally, the solutions $\psi^{\left(  W\right)  }$ (\ref{DiracW}) and
$\psi^{\left(  DP\right)  }$ (\ref{DiracDP}) are represented by $4\times2$
matrices whose second columns are zero and whose first columns give the usual
Dirac bispinors of quantum theory. If $R$ is replaced by the de Broglie spin
rotor $R=\exp\left[  -i\mathbf{e}_{3}\left\langle p\bar{x}\right\rangle
_{S}/\hbar\right]  ,$ the solutions in the nonvanishing columns of
$\psi^{\left(  W\right)  }$ and $\psi^{\left(  DP\right)  }$ are the usual
momentum eigenstates of the Dirac equation.

To get the Dirac equation in the Dirac-Pauli representation, we split
$p\bar{\Psi}^{\dag}=m\Psi$ into even and odd parts:%
\begin{align}
-\mathbf{p}\left\langle \Psi\right\rangle _{-}  &  =\left(  m-p^{0}\right)
\left\langle \Psi\right\rangle _{+}\\
\mathbf{p}\left\langle \Psi\right\rangle _{+}  &  =\left(  m+p^{0}\right)
\left\langle \Psi\right\rangle _{-}~.
\end{align}
The odd part%
\begin{equation}
\left\langle \Psi\right\rangle _{-}=\left(  m+p^{0}\right)  ^{-1}%
\mathbf{p}\left\langle \Psi\right\rangle _{+}%
\end{equation}
can be eliminated in the first equation to give a second-order form of the
Dirac equation:%
\begin{equation}
\mathbf{p}\left(  m+p^{0}\right)  ^{-1}\mathbf{p}\left\langle \Psi
\right\rangle _{+}=\left(  p^{0}-m\right)  \left\langle \Psi\right\rangle
_{+}. \label{Dirac2}%
\end{equation}

\subsection{Differential Operators and Commutation Relations}

The differential form of momentum results when we (1) assume that the general
$\Psi$ can be expressed as a linear superposition of de Broglie waves
$R~\exp\left(  -i\left\langle x\bar{p}\right\rangle _{S}\mathbf{e}_{3}%
/\hbar\right)  $ and (2) make the usual local gauge invariance argument for
the spacetime vector potential $A.$ In APS we can write for each component%
\begin{equation}
p^{\mu}\Psi=i\hbar\partial^{\mu}\Psi\mathbf{e}_{3}-eA^{\mu}\Psi. \label{p_mu}%
\end{equation}
When used with the classical Dirac equation, the result is fully equivalent to
the usual Dirac equation in its differential form. When the $\mu=0$ equation
(\ref{p_mu}) is applied to the even part of $\Psi$ and projected onto the
minimal left ideal $\left(  \text{APS}\right)  $P$_{3},$ we find%
\begin{equation}
p^{0}\left\langle \Psi\right\rangle _{+}\text{P}_{3}=\left(  i\hbar
\partial_{t}-V\right)  \left\langle \Psi\right\rangle _{+}\text{P}_{3}=\left(
H-V\right)  \left\langle \Psi\right\rangle _{+}\text{P}_{3}~,
\end{equation}
where the Hamiltonian operator $H=i\hbar\partial_{t}$ operating on
$\left\langle \Psi\right\rangle _{+}$P$_{3}$ includes the rest energy $m.$ The
projected form of the second-order Dirac equation (\ref{Dirac2}) thus becomes%
\begin{equation}
\mathbf{p}\left(  m+p^{0}\right)  ^{-1}\mathbf{p}\left\langle \Psi
\right\rangle _{+}\text{P}_{3}=\left(  H-V-m\right)  \left\langle
\Psi\right\rangle _{+}\text{P}_{3} \label{Dirac2a}%
\end{equation}
In the low-energy limit where the factor $\left(  m+p^{0}\right)  ^{-1}$ on
the LHS is approximated by $\left(  2m\right)  ^{-1},$ Eq. (\ref{Dirac2a})
becomes the Pauli-Schr\"{o}dinger equation. This demonstrates that the
Pauli-Schr\"{o}dinger wave function (with a two-component spinor of the form
(\ref{RP_3}) corresponds to the even ideal spinor $\left\langle \Psi
\right\rangle _{+}$P$_{3}$ in the low-energy limit (\ref{Even_Psi}):%
\begin{equation}
\psi^{\left(  \text{Sch}\right)  }=\rho^{1/2}R\text{P}_{3}~. \label{Psi_Sch}%
\end{equation}

The differential-operator form of $p^{\mu}$ (\ref{p_mu}) implies the
commutation relations%
\begin{equation}
\left[  p^{\mu},x_{\nu}\right]  \Psi=i\hbar\delta_{\nu}^{\mu}\Psi
\mathbf{e}_{3} \label{comm_rel1}%
\end{equation}
In terms of the spin biparavector $\mathbf{S}=\Lambda e_{1}\bar{e}_{2}%
\bar{\Lambda}=-i\Lambda e_{3}\bar{\Lambda}$ (\ref{PLdual}), we note%
\begin{equation}
i\Psi\mathbf{e}_{3}=-\mathbf{S}\Psi
\end{equation}
so that relation (\ref{comm_rel1}) is equivalent to $\left[  p^{\mu},x_{\nu
}\right]  \Psi=-\hbar\mathbf{S}\Psi,$ and since this is true for any current
amplitude $\Psi,$ we can simply write the operator relation%
\begin{equation}
\left[  p^{\mu},x_{\nu}\right]  =-\hbar\delta_{\nu}^{\mu}\mathbf{S.}
\label{comm_rel2}%
\end{equation}
In the minimal left ideal $\left(  \text{APS}\right)  $P$_{3},$ the relation
(\ref{comm_rel2}) reduces to the usual form, in which $\mathbf{S}$ is replaced
by $-i.$

We have treated the momentum components $p^{\mu}$ as operators and the
coordinates $x^{\mu}$ as variables, but the inherent symmetry of momentum and
position variables as apparent in phase-space treatments and in Eq.
(\ref{comm_rel2}) allows us to reverse the roles and consider $p^{\mu}$ the
coordinates in momentum space and to write the $x^{\mu}$ as differential
operators on this space. However, there is a well-known objection by
Pauli\cite{Pauli58} to considering time as an operator satisfying $\left[
H,t\right]  $ $=i\hbar,$ where we identify $p^{0}=H,\ x_{0}=t.$ He pointed out
that given any energy eigenstate $\psi_{E}$ with eigenvalue $E,$ $H\psi
_{E}=E\psi_{E},$ one could then form another eigenstate $\psi_{E-\varepsilon
}=\exp\left(  i\varepsilon t/\hbar\right)  \psi_{E}$ of eigenenergy
$E-\varepsilon,$ since $H\exp\left(  i\varepsilon t/\hbar\right)  =\exp\left(
i\varepsilon t/\hbar\right)  \left(  H+i\varepsilon\left[  H,t\right]
/\hbar\right)  =\exp\left(  i\varepsilon t/\hbar\right)  \left(
H-\varepsilon\right)  .$ The spectrum of $H$ must then be continuous and
unbounded. In the calculation (\ref{magdipmmt}) of the magnetic dipole moment
of a fermion, we have shown that the \textquotedblleft Pauli
problem\textquotedblright\ is avoided in APS. The phase factor $\exp\left(
i\varepsilon t/\hbar\right)  $ arises only from the projection of an actual
rotation such as caused by a magnetic field, and such a constant rotation rate
does indeed change the energy. It is that change in energy that gave us the
correct magnetic moment and that can supply the energy when a spin is
accelerated in a magnetic-field gradient.

Note that our approach also gives a natural, relativistic formulation of the
Bohm/de Broglie theory\cite{deBroglie85,deBroglie64,Bohm93} of causal quantum
mechanics with spin included. Flow lines are given by the current density
$J=\Psi e_{0}\Psi^{\dag},$ and the Pauli-Luba\'{n}ski spin distribution is
found from $\mathfrak{S}=\Psi e_{3}\Psi^{\dag}.$ Of course, the existence of a
simple formalism does not imply that Bohm's causal interpretation is required.

The combined currents $\mathfrak{J}_{\pm}=J\pm\mathfrak{S}$ are null elements,
for which continuity equations are readily established:%
\begin{equation}
\left\langle \bar{\partial}\mathfrak{J}_{\pm}\right\rangle _{S}=\left\langle
\bar{\partial}\left[  \Psi\left(  e_{0}\pm e_{3}\right)  \Psi^{\dag}\right]
\right\rangle _{S}=2\left\langle \left(  \bar{\partial}\Psi\right)  \left(
e_{0}\pm e_{3}\right)  \Psi^{\dag}\right\rangle _{\Re S}%
\end{equation}
since from the Dirac equation (\ref{class_Dirac}) with (\ref{p_mu}), the
relation%
\begin{equation}
\bar{\partial}\Psi=-i\hbar^{-1}\left(  \bar{p}+e\bar{A}\right)  \Psi
e_{3}=-i\hbar^{-1}\left(  m\bar{\Psi}^{\dag}+e\bar{A}\Psi\right)  e_{3}%
\end{equation}
gives%
\begin{align}
\left\langle \bar{\partial}\mathfrak{J}_{\pm}\right\rangle _{S}  &
=-2\hbar^{-1}\left\langle i\left(  \left(  m\bar{\Psi}^{\dag}+e\bar{A}%
\Psi\right)  e_{3}\right)  \left(  e_{0}\pm e_{3}\right)  \Psi^{\dag
}\right\rangle _{\Re S}\\
&  =-2\hbar^{-1}\left\langle i\left(  \left(  m\rho+e\Psi^{\dag}\bar{A}%
\Psi\right)  \right)  \left(  e_{3}\pm e_{0}\right)  \right\rangle _{\Re
S}=0.\nonumber
\end{align}
This means that both $\mathfrak{J}_{\pm}$ are conserved currents, and
therefore, so are $J$ and $\mathfrak{S.}$ The other Fierz identities for
bilinear covariants also follow.\cite{Bay96}

\section{Spin Distributions}

Once the momentum has been replaced by a differential operator and the spatial
form of the Dirac equation has been derived, we have crossed into the quantum
side of the quantum/classical interface. While we needed to establish the
correspondence between the Dirac spinor and Pauli-Schr\"{o}dinger wave
function to our classical eigenspinor, we consider here a more classical
calculation of spin distributions. We want to show that the calculation is
exactly equivalent to the corresponding quantum-mechanical one. To study spin
distributions, the low-velocity limit
\begin{equation}
\Psi\simeq\left\langle \Psi\right\rangle _{+}\simeq\rho^{1/2}R \label{eslowv}%
\end{equation}
is sufficient, with $R$ given in terms of Euler angles by (\ref{Euler1}). As
we saw above (\ref{Psi_Sch}), the Schr\"{o}dinger wave function $\psi^{\left(
\text{Sch}\right)  }$ with a two-component spinor, is the projection of this
$\Psi$ (\ref{eslowv}) onto the minimal left ideal $\left(  \text{APS}\right)
$P$_{3}.$ To get expressions at relativistic speeds, we can always apply a
subsequent boost to $\Psi$. The classical spin direction in a static system is
$\mathbf{s=}R\mathbf{e}_{3}R^{\dag}$ A Pauli-Luba\'{n}ski spin distribution of
the state is
\begin{equation}
\mathfrak{S}=\rho~\mathbf{s}=\rho R\mathbf{e}_{3}R^{\dag}=\Psi\mathbf{e}%
_{3}\Psi^{\dag},
\end{equation}
where the positive scalar $\rho=\rho\left(  \mathbf{r}\right)  $ is the
density of spins in the reference frame. As seen below, simple measurements of
the spin direction give only one component at a time. The distribution of the
component of the spin in the direction of an arbitrary unit vector
$\mathbf{m}$ is%
\begin{equation}
\rho~\mathbf{s\cdot m=}\left\langle \mathfrak{S}\mathbf{m}\right\rangle
_{S}=\left\langle \Psi\mathbf{e}_{3}\Psi^{\dag}\mathbf{m}\right\rangle _{S}\,.
\end{equation}
In terms of the projector $\mathsf{P}_{3}=$ P$_{3}^{2},$ since $\mathbf{e}%
_{3}=\mathsf{P}_{3}-\mathsf{\bar{P}}_{3},$ $\Psi$ is even, and for any
elements $p,q,$ $\left\langle pq\right\rangle _{S}=\left\langle
qp\right\rangle _{S}=\left\langle \overline{pq}\right\rangle _{S}\,$, the
distribution is
\begin{align}
\left\langle \Psi\mathsf{P}_{3}\Psi^{\dag}\mathbf{m}\right\rangle
_{S}-\left\langle \Psi\mathsf{\bar{P}}_{3}\Psi^{\dag}\mathbf{m}\right\rangle
_{S}  &  =\left\langle \Psi\mathsf{P}_{3}\Psi^{\dag}\mathbf{m}\right\rangle
_{S}+\left\langle \mathbf{m}\bar{\Psi}^{\dag}\mathsf{P}_{3}\bar{\Psi
}\right\rangle _{S}\\
&  =2\left\langle \Psi\mathsf{P}_{3}\Psi^{\dag}\mathbf{m}\right\rangle
_{S}=2\rho\left\langle R\text{P}_{3}R^{\dag}\mathbf{m}\right\rangle
_{S}\label{spinstate}\\
&  =2\left\langle \mathsf{P}_{3}\Psi^{\dag}\mathbf{m}\Psi\mathsf{P}%
_{3}\right\rangle _{S}=\mathrm{tr}\left\{  \psi^{\left(  \text{Sch}\right)
\dag}\mathbf{m}\psi^{\left(  \text{Sch}\right)  }\right\}  \,,
\label{spindistr}%
\end{align}
where $\psi^{\left(  \text{Sch}\right)  }$ has the standard matrix
representation%
\begin{equation}
\psi^{\left(  \text{Sch}\right)  }\equiv\rho^{1/2}R\mathsf{P}_{+3}%
=e^{-i\chi/2}\rho^{1/2}\left(
\begin{array}
[c]{cc}%
e^{-i\phi/2}\cos\theta/2 & 0\\
e^{i\phi/2}\sin\theta/2 & 0
\end{array}
\right)  \label{psi_Schr}%
\end{equation}
which, ignoring the inconsequential column of zeros, is the two-component
spinor familiar from the usual nonrelativistic Pauli theory. The term
$\psi^{\left(  \text{Sch}\right)  \dag}\mathbf{m}\psi^{\left(  \text{Sch}%
\right)  }$ is then a scalar and $\mathrm{tr}$ can be omitted from
(\ref{spindistr}). If $\rho$ is normalized to unity,%
\begin{equation}
\int d^{3}\mathbf{x}~\rho=2\int d^{3}\mathbf{x}~\left\langle \psi^{\left(
\text{Sch}\right)  \dag}\psi^{\left(  \text{Sch}\right)  }\right\rangle
_{S}\equiv\left\langle \psi^{\left(  \text{Sch}\right)  }|\psi^{\left(
\text{Sch}\right)  }\right\rangle =1,
\end{equation}
then the average component of the spin in the direction $\mathbf{m}$ is%
\begin{equation}
2\int d^{3}\mathbf{x}~\left\langle \psi^{\left(  \text{Sch}\right)  \dag
}\mathbf{m}\psi^{\left(  \text{Sch}\right)  }\right\rangle _{S}\equiv
\left\langle \psi^{\left(  \text{Sch}\right)  }\left\vert \mathbf{m}%
\right\vert \psi^{\left(  \text{Sch}\right)  }\right\rangle .
\end{equation}

Although we derived the spin distribution as a classical expression, it has
\emph{precisely the quantum form }if we recognize that the matrix
representation of the unit vector $\mathbf{m}$, namely $\mathbf{m}%
=m^{j}\mathbf{e}_{j}\rightarrow m^{1}\sigma_{x}+m^{2}\sigma_{y}+m^{3}%
\sigma_{z}\,,$ is traditionally written $\mathbf{m\cdot}\boldsymbol{\sigma}$
(but this is misleading, since it represents a vector, not a scalar) and
traditionally, $\boldsymbol{\sigma}$ is thought of as the spin operator.

From expression (\ref{spinstate}) we see that the real paravector
$\mathsf{P}_{\mathbf{s}}=R\mathsf{P}_{+3}R^{\dag}=\frac{1}{2}\left(
1+\mathbf{s}\right)  $ embodies information about the classical spin state.
Then P$_{\mathbf{s}}=\psi^{\left(  \text{Sch}\right)  }\psi^{\left(
\text{Sch}\right)  \dag}$ is the spin density operator $\varrho$ for the pure
state $\psi^{\left(  \text{Sch}\right)  }.$of spin $\mathbf{s.}$ It is also a
projector that acts as a state filter. To see whether a system with spin
density $\varrho$ is in a given state of spin $\mathbf{n}$ we can apply the
state filter to the spin density operator $\varrho$ and see what remains:%
\begin{equation}
\mathsf{P}_{\mathbf{n}}\varrho\mathsf{P}_{\mathbf{n}}=\left(  \mathsf{P}%
_{\mathbf{n}}\varrho+\bar{\varrho}\mathsf{\bar{P}}_{\mathbf{n}}\right)
\mathsf{P}_{\mathbf{n}}=2\left\langle \mathsf{P}_{\mathbf{n}}\varrho
\right\rangle _{S}\mathsf{P}_{\mathbf{n}}\,. \label{filter}%
\end{equation}
The scalar coefficient $2\left\langle \mathsf{P}_{\mathbf{n}}\varrho
\right\rangle _{S}=\left\langle \left(  1+\mathbf{n}\right)  \varrho
\right\rangle _{S}$ is the probability of finding the system described by
$\varrho$ in the state $\mathbf{n\,.}$ For a system in the pure state
$\varrho=\mathsf{P}_{\mathbf{s}}=\frac{1}{2}\left(  1+\mathbf{s}\right)  ,$
the probability is
\begin{equation}
2\left\langle \mathsf{P}_{\mathbf{n}}\mathsf{P}_{\mathbf{s}}\right\rangle
_{S}=\frac{1}{2}\left\langle \left(  1+\mathbf{n}\right)  \left(
1+\mathbf{s}\right)  \right\rangle _{S}=\frac{1}{2}\left(  1+\mathbf{n\cdot
s}\right)  . \label{prob}%
\end{equation}
This is unity if the system is definitely in the state $\mathbf{n,}$ whereas
it vanishes if the system is in a state \emph{orthogonal} to $\mathbf{n\,.}$
Thus, $\mathbf{s}=\mathbf{n}$ is required for the states to be the same and
$\mathbf{s}=-\mathbf{n}$ for the states to be orthogonal. Note that the
mathematics is the same as used to describe light
polarization.\cite{Bay04,Bay93}

\subsection{Spin $\frac{1}{2}$ and State Expansions}

The value of $\frac{1}{2}$ for the spin of elementary spinors in physical
space can be associated with the group-theoretical label for the irreducible
spinor representation of the rotation group $SU\left(  2\right)  $ carried by
ideal spinors, but it is also required by the fact that any rotation can be
expressed as a linear superposition of two orthogonal rotations defined for
any direction in space. The Euler-angle form (\ref{Euler1}) of any rotor $R$
can be rewritten
\begin{align}
R &  =\exp\left(  -i\mathbf{n}\theta/2\right)  \exp\left[  -i\mathbf{e}%
_{3}\left(  \phi+\chi\right)  /2\right]  \,\label{Euler}\\
&  =\left(  \cos\frac{\theta}{2}-i\mathbf{n}\sin\frac{\theta}{2}\right)
\exp\left[  -i\mathbf{e}_{3}\left(  \phi+\chi\right)  /2\right]  ,\nonumber
\end{align}
where $\mathbf{n=}\exp\left(  -i\mathbf{e}_{3}\phi/2\right)  \mathbf{e}%
_{2}\exp\left(  i\mathbf{e}_{3}\phi/2\right)  $ is a unit vector in the
$\mathbf{e}_{1}\mathbf{e}_{2}$ plane. Therefore, any rotor $R$ is a real
linear combination $\cos\frac{\theta}{2}R_{\uparrow}+\sin\frac{\theta}%
{2}R_{\downarrow}$ of rotors $R_{\uparrow}=\exp\left[  -i\mathbf{e}_{3}\left(
\phi+\chi\right)  /2\right]  $ and $R_{\downarrow}=-i\mathbf{n}R_{\uparrow}$
that are mutually orthogonal: $\left\langle R_{\uparrow}\bar{R}_{\downarrow
}\right\rangle _{S}=\left\langle R_{\uparrow}R_{\downarrow}^{\dag
}\right\rangle _{S}=\left\langle -i\mathbf{n}\right\rangle _{S}=0~.$ The rotor
$R_{\uparrow}$ maintains the $\mathbf{e}_{3}$ component whereas $R_{\downarrow
}$ flips it.

By projecting the rotors with P$_{3}$ onto the corresponding minimal left
ideal, we obtain the equivalent relation of ideal spinors, which represent
states with a given spin orientation:
\begin{subequations}
\begin{align}
\psi &  =R\text{P}_{3}=\left(  \cos\frac{\theta}{2}R_{\uparrow}+\sin
\frac{\theta}{2}R_{\downarrow}\right)  \text{P}_{3}\nonumber\\
&  =\cos\frac{\theta}{2}\psi_{\uparrow}+\sin\frac{\theta}{2}\psi_{\downarrow
}\label{idealspin}\\
\psi_{\uparrow}  &  =e^{-i\left(  \phi+\chi\right)  /2}\text{P}_{3}%
,\;\psi_{\downarrow}=-i\mathbf{n}\psi_{\uparrow}~.\nonumber
\end{align}
The projection operators P$_{3}$ and \={P}$_{3}$ operating from the left
isolate the spin-up and spin-down parts:
\end{subequations}
\begin{align}
\text{P}_{3}\psi &  =\cos\frac{\theta}{2}\psi_{\uparrow}\label{updown}\\
\text{\={P}}_{3}\psi &  =\sin\frac{\theta}{2}\psi_{\downarrow}%
\end{align}
both of which are eigenstates of $\mathbf{e}_{3}:$%
\begin{align}
\mathbf{e}_{3}\text{P}_{3}\psi &  =+\text{P}_{3}\psi\\
\mathbf{e}_{3}\text{\={P}}_{3}\psi &  =-\text{\={P}}_{3}\psi.
\end{align}
Traditional orthonormality conditions hold:
\begin{align}
2\left\langle \psi_{\uparrow}\psi_{\downarrow}^{\dag}\right\rangle _{S}  &
=2\left\langle \psi_{\uparrow}\psi_{\uparrow}^{\dag}i\mathbf{n}\right\rangle
_{S}=2\rho\left\langle \text{P}_{+3}i\mathbf{n}\right\rangle _{S}=0\\
2\left\langle \psi_{\downarrow}\psi_{\downarrow}^{\dag}\right\rangle _{S}  &
=2\left\langle \psi_{\uparrow}\psi_{\uparrow}^{\dag}\right\rangle _{S}%
=2\rho\left\langle \text{P}_{+3}\right\rangle _{S}=\rho~.
\end{align}
It follows that the amplitudes are%
\begin{align}
\left\langle \psi_{\uparrow}|\psi\right\rangle  &  =2\left\langle \psi
\psi_{\uparrow}^{\dag}\right\rangle _{S}=\cos\frac{\theta}{2}\\
\left\langle \psi_{\downarrow}|\psi\right\rangle  &  =2\left\langle \psi
\psi_{\downarrow}^{\dag}\right\rangle _{S}=\sin\frac{\theta}{2}%
\end{align}
giving probabilities as found above in Eq. (\ref{prob}).%
\begin{align}
\left\vert \left\langle \psi_{\uparrow}|\psi\right\rangle \right\vert ^{2}  &
=\cos^{2}\frac{\theta}{2}=\frac{1}{2}\left(  1+\mathbf{s\cdot e}_{3}\right)
\label{ampl1}\\
\left\vert \left\langle \psi_{\downarrow}|\psi\right\rangle \right\vert ^{2}
&  =\sin^{2}\frac{\theta}{2}=\frac{1}{2}\left(  1-\mathbf{s\cdot e}%
_{3}\right)  . \label{ampl2}%
\end{align}

\subsection{Stern-Gerlach Experiment}%

\index{Stern-Gerlach experiment}%
The basic measurement of spin is that of the Stern-Gerlach
experiment\cite{Ball98}, in which a beam of ground-state silver atoms is split
by a magnetic-field gradient into distinct beams of opposite spin
polarization. It is a building block of real and thought experiments in
quantum measurement\cite{Alb94}. A description succeeds in the classical
eigenspinor framework because of the linear form of the equations, the
consequent possibility of superposition, and explicitly the ability to write
rotors as superpositions of \textquotedblleft spin-up\textquotedblright\ and
\textquotedblleft spin-down\textquotedblright\ rotors referenced to any direction.

Consider a nonrelativistic beam of ground-state atoms that travels with
velocity $\mathbf{v}=v\mathbf{e}_{1}$ through a static magnetic field
$\mathbf{B}$ that vanishes everywhere except in the vicinity of the
Stern-Gerlach magnet, where it has a gradient aligned with the $\mathbf{e}%
_{3}$ ($z$) axis. The net effect of the magnetic gradient on atoms in the beam
is to apply an impulse or boost in the $\mathbf{e}_{3}$ direction proportional
to the $z$ component $\mu_{z}$ of the magnetic dipole moment.

The generic form of the Stern-Gerlach experiment can be put into a form
analogous to the action of a birefringent crystal on a beam of polarized
light: the ideal state spinor (\ref{idealspin}) is split into two parts%
\begin{equation}
\psi=\left(  \text{P}_{3}+\text{\={P}}_{3}\right)  \psi
\end{equation}
which become separated spatially. In the Stern-Gerlach case, each part is
associated with a distinct boost, so that the full state spinor $\Psi$ becomes%
\begin{equation}
\Psi=2B_{+}\left\langle \text{P}_{3}\psi\right\rangle _{+}+2B_{-}\left\langle
\text{\={P}}_{3}\psi\right\rangle _{+},\label{SGPsi}%
\end{equation}
where the boosts [see Eq. (\ref{B_even_odd})] combine the velocity
$\mathbf{v}=v\mathbf{e}_{1}$ of the beam before the magnetic-field gradient
with increments $\pm\Delta v\mathbf{e}_{3}$ induced by the field gradient. In
the nonrelativistic limit%
\begin{equation}
B_{\pm}\simeq1+\frac{1}{2}\left(  v\mathbf{e}_{1}\pm\Delta v\mathbf{e}%
_{3}\right)  \equiv1+\frac{1}{2}\mathbf{V}_{\pm}.
\end{equation}
With $\phi=0,$ $\chi=\left\langle p\bar{x}\right\rangle _{S},\ $and
$\mathbf{n}=\mathbf{e}_{2},$ we have $\psi_{\uparrow}=\rho^{1/2}e^{-i\left(
\phi+\chi\right)  /2}$P$_{3},\;\psi_{\downarrow}=-i\mathbf{n}\psi_{\uparrow}~$%
\begin{align}
\text{P}_{3}\psi &  =\cos\frac{\theta}{2}\psi_{\uparrow}=\rho^{1/2}%
e^{-i\chi/2}\cos\frac{\theta}{2}\text{P}_{3}\nonumber\\
2\left\langle \text{P}_{3}\psi\right\rangle _{+} &  =\rho^{1/2}e^{-i\chi
\mathbf{e}_{3}/2}\cos\frac{\theta}{2}\\
\text{\={P}}_{3}\psi &  =\sin\frac{\theta}{2}\psi_{\downarrow}=\rho
^{1/2}\mathbf{e}_{1}e^{-i\chi\mathbf{e}_{3}/2}\sin\frac{\theta}{2}\text{P}%
_{3}\nonumber\\
2\left\langle \text{\={P}}_{3}\psi\right\rangle _{+} &  =\rho^{1/2}%
\mathbf{e}_{1}\mathbf{e}_{3}e^{-i\chi\mathbf{e}_{3}/2}\sin\frac{\theta}{2}~.
\end{align}
Thus, $\Psi$ (\ref{SGPsi}) becomes%
\begin{equation}
\Psi=\rho^{1/2}\left(  B_{+}\cos\frac{\theta}{2}+B_{-}\sin\frac{\theta}%
{2}\mathbf{e}_{1}\mathbf{e}_{3}\right)  e^{-i\chi\mathbf{e}_{3}/2}.
\end{equation}

If the initial beam has finite profile $\rho\left(  x\right)  $, the action of
the Stern-Gerlach magnet will eventually split the beam into two distinct
beams moving with velocities $\mathbf{V}_{\pm}$ and with opposite spin
directions and distinct profiles $\rho_{\pm}\simeq\rho\left(  x-\mathbf{V}%
_{\pm}t\right)  $. The proper-velocity profile is given by the current density%
\begin{align}
J &  =\Psi e_{0}\Psi^{\dag}=\left(  B_{+}\cos\frac{\theta}{2}+B_{-}\sin
\frac{\theta}{2}\mathbf{e}_{1}\mathbf{e}_{3}\right)  \rho\left(  B_{+}%
\cos\frac{\theta}{2}+\sin\frac{\theta}{2}\mathbf{e}_{3}\mathbf{e}_{1}%
B_{-}\right)  \nonumber\\
&  =B_{+}^{2}\rho_{+}\cos^{2}\frac{\theta}{2}+B_{-}^{2}\rho_{-}\sin^{2}%
\frac{\theta}{2}+\rho\sin\theta\left\langle B_{+}\mathbf{e}_{3}\mathbf{e}%
_{1}B_{-}\right\rangle _{\Re}%
\end{align}
since the cross terms cancel. At some distance down the beam past the magnet,
the 2 sub-beams become non-overlapping and there are two distinct beams. The
corresponding spin-density profile is%
\begin{align}
\mathfrak{S} &  =\Psi\mathbf{e}_{3}\Psi^{\dag}=\left(  B_{+}\cos\frac{\theta
}{2}+B_{-}\sin\frac{\theta}{2}\mathbf{e}_{1}\mathbf{e}_{3}\right)
\mathbf{e}_{3}\rho\left(  B_{+}\cos\frac{\theta}{2}+\sin\frac{\theta}%
{2}\mathbf{e}_{3}\mathbf{e}_{1}B_{-}\right)  \nonumber\\
&  =\rho_{+}B_{+}\mathbf{e}_{3}B_{+}\cos^{2}\frac{\theta}{2}-\rho_{-}%
B_{-}\mathbf{e}_{3}B_{-}\sin^{2}\frac{\theta}{2}+\rho\sin\theta\left\langle
B_{+}\mathbf{e}_{1}B_{-}\right\rangle _{\Re}.
\end{align}
Algebraically, the functions $\rho_{\pm}$ are the same as $\rho$ but the
argument is appropriately transformed from the rest-frame coordinates. For
consistency, we can take $B\rho^{\frac{1}{2}}=\left(  B\rho B\right)  ^{1/2}$
so that the $\rho$ factors in the $\sin\theta$ terms in the above expressions
for $J$ and $\mathfrak{S}$ are to be calculated as the geometric mean $\left(
\rho_{+}\rho_{-}\right)  ^{1/2}.$ Using $B_{\pm}=1+\frac{1}{2}\mathbf{V}_{\pm
}$ and discarding terms quadratic in $\mathbf{V}_{\pm},$ we get%
\begin{align}
B_{\pm}^{2} &  \simeq1+\mathbf{V}_{\pm}\nonumber\\
B_{+}\mathbf{e}_{3}B_{+} &  =\mathbf{e}_{3}+2\mathbf{V}_{+}\cdot\mathbf{e}%
_{3}=\mathbf{e}_{3}+\Delta v\nonumber\\
B_{-}\mathbf{e}_{3}B_{-} &  =\mathbf{e}_{3}-\Delta v\\
\left\langle B_{+}\mathbf{e}_{3}\mathbf{e}_{1}B_{-}\right\rangle _{\Re} &
=\frac{1}{2}\left(  \mathbf{V}_{+}-\mathbf{V}_{-}\right)  \mathbf{e}%
_{3}\mathbf{e}_{1}=\Delta v\mathbf{e}_{1}\nonumber\\
\left\langle B_{+}\mathbf{e}_{1}B_{-}\right\rangle _{\Re} &  =\mathbf{e}%
_{1}+v\nonumber
\end{align}
This gives%
\begin{align}
J &  =\frac{1}{2}\rho_{+}\left(  1+\mathbf{V}_{+}\right)  \left(  1+\cos
\theta\right)  +\frac{1}{2}\rho_{-}\left(  1+\mathbf{V}_{-}\right)  \left(
1-\cos\theta\right)  +\sqrt{\rho_{+}\rho_{-}}\Delta v\mathbf{e}_{1}\sin
\theta\\
\mathfrak{S} &  =\frac{1}{2}\rho_{+}\left(  \mathbf{e}_{3}+\Delta v\right)
\left(  1+\cos\theta\right)  -\frac{1}{2}\rho_{-}\left(  \mathbf{e}_{3}-\Delta
v\right)  \left(  1-\cos\theta\right)  +\sqrt{\rho_{+}\rho_{-}}\sin
\theta\left(  \mathbf{e}_{1}+v\right)  .
\end{align}
The last term in each expression is an interference contribution that dies
away as the distributions $\rho_{\pm}$ separate and cease to overlap. The
scalar terms in the spin distribution arise because of relativity: it is
really a distribution of the Pauli-Luba\'{n}ski spin, and the boost of the
rest-frame spin has a scalar contribution, but this is small because the
velocities involved are much less than 1 (the speed of light).

\subsection{Uncertainty in Spin Measurements}

We identified above the quantum spin operators usually denoted
$\boldsymbol{\sigma}\cdot\mathbf{n,}$ with unit vectors $\mathbf{n}$ in APS.
Since the commutation relations of the operators are the same as those for the
vectors, it is not too surprising that \textquotedblleft uncertainty
relations\textquotedblright\ exist among the measured classical spin
components. Nevertheless,it is worthwhile to display the relation explicitly
as it bears on our interpretation of spin. From the commutation relation
$\left[  \sigma_{x},\sigma_{y}\right]  =2i\sigma_{z},$ one can derive the
uncertainty relation\cite{Bal96}%
\begin{equation}
\Delta\sigma_{x}\Delta\sigma_{y}\geq\left\vert \left\langle \sigma
_{z}\right\rangle \right\vert .\label{spin_uncertain}%
\end{equation}
Analogous relations hold for cyclic permutations $x\rightarrow y\rightarrow
z\rightarrow x$ of the indices. The relation holds classically as may be seen
by a couple of examples.

For a system with classical spin $\mathbf{s}=\mathbf{e}_{3},$ we can write the
rotor $R=1$ as the superposition of rotors%
\begin{equation}
R=\frac{1}{\sqrt{2}}\left[  \exp\left(  \mathbf{e}_{1}\mathbf{e}_{3}%
\pi/4\right)  +\exp\left(  -\mathbf{e}_{1}\mathbf{e}_{3}\pi/4\right)  \right]
~,
\end{equation}
which expands the state%
\begin{align}
R\text{P}_{3} &  =\left[  \frac{1}{2}\left(  1+\mathbf{e}_{1}\mathbf{e}%
_{3}\right)  +\frac{1}{2}\left(  1-\mathbf{e}_{1}\mathbf{e}_{3}\right)
\right]  \text{P}_{3}\\
&  =\text{P}_{1}\text{P}_{3}+\text{\={P}}_{1}\text{P}_{3}\nonumber
\end{align}
into eigenstates of $\mathbf{e}_{1}$ with equal portions filtered in the
$+\mathbf{e}_{1}$ and $-\mathbf{e}_{1}$ directions. It follows for
measurements of $\mathbf{s}$ along $\mathbf{e}_{1}$ that the root-mean-square
deviation $\Delta\sigma_{x}$ is $+1.$ The argument is similar for
$\Delta\sigma_{y}$ except that we split $R$ into equal parts of $\exp\left(
\pm\mathbf{e}_{2}\mathbf{e}_{3}\pi/4\right)  .$ We thus find $\Delta\sigma
_{x}\Delta\sigma_{y}=1=\left\langle \sigma_{z}\right\rangle .$ On the other
had, if we start in an eigenstate of $\mathbf{e}_{1},$ such as $R$P$_{3}%
=\exp\left(  \mathbf{e}_{1}\mathbf{e}_{3}\pi/4\right)  $P$_{3}=\sqrt{2}$%
P$_{1}$P$_{3}$ we find $\Delta\sigma_{x}=0=\left\langle \sigma_{z}%
\right\rangle ,$ so that once again the uncertainty relation is satisfied.

More generally, any state $\psi=R$P$_{3}$ has the probability $2\left\langle
\psi\psi^{\dag}\text{P}_{\mathbf{n}}\right\rangle _{S}=2\left\langle
\text{P}_{\mathbf{s}}\text{P}_{\mathbf{n}}\right\rangle _{S}=\frac{1}%
{2}\left(  1+\mathbf{s\cdot n}\right)  $ of being measured with spin
$\mathbf{n,}$ where P$_{\mathbf{s}}=R$P$_{3}R^{\dag}.$ The average spin after
filtering with P$_{\pm\mathbf{n}}$ is therefore $\frac{1}{2}\left(
1+\mathbf{s\cdot n}\right)  -\frac{1}{2}\left(  1-\mathbf{s\cdot n}\right)
=\mathbf{s\cdot n,}$ and the mean square deviation of the measurement is%
\begin{align}
\left(  \Delta\mathbf{s\cdot n}\right)  ^{2} &  =\frac{1}{2}\left(
1+\mathbf{s\cdot n}\right)  \left(  1-\mathbf{s\cdot n}\right)  ^{2}+\frac
{1}{2}\left(  1-\mathbf{s\cdot n}\right)  \left(  1+\mathbf{s\cdot n}\right)
^{2}\\
&  =1-\left(  \mathbf{s\cdot n}\right)  ^{2}\nonumber
\end{align}
The product $\Delta\mathbf{s\cdot e}_{1}\Delta\mathbf{s\cdot e}_{2}$ of
root-mean-square deviations is thus%
\begin{align}
\Delta\mathbf{s\cdot e}_{1}\Delta\mathbf{s\cdot e}_{2} &  =\sqrt{\left(
1-\left(  \mathbf{s\cdot e}_{1}\right)  ^{2}\right)  \left(  1-\left(
\mathbf{s\cdot e}_{2}\right)  ^{2}\right)  }\\
&  \geq\sqrt{1-\left(  \mathbf{s\cdot e}_{1}\right)  ^{2}-\left(
\mathbf{s\cdot e}_{2}\right)  ^{2}}=\left\vert \mathbf{s\cdot e}%
_{3}\right\vert \nonumber
\end{align}
which is equivalent to the uncertainty relation (\ref{spin_uncertain})

\section{Conclusions}

Classical origins of fermionic spin 1/2 in the geometry of physical space are
suggested by the classical formulation of elementary-particle dynamics in APS.
APS itself, the geometric algebra of physical space, can be generated from a
pair of fermion annihilation and creation operators, or equivalently from
spin-1/2 raising and lowering operators. Spatial rotors, when projected onto a
minimal left ideal of APS, are identified as two-component Pauli spinors, and
experience the same change of sign under 360-degree rotations. This
association is extended in a relativistic treatment to a correspondence
between the classical eigenspinor $\Lambda,$ the Lorentz amplitude for
transformations between the rest and lab frames, and the four-component
quantum Dirac spinor $\Psi.$ The eigenspinor gives the orientation and motion
of the particle frame as seen in the lab, and from it, an amplitude of the
current density is formed, which satisfies linear equations and allows for
superposition and quantum-like interference. A simple derivation of the
classical Dirac equation, and its close relation to quantum formalism,
illuminates the Q/C interface and demonstrates that many quantum phenomena
have classical roots..Relativity is an essential part of this approach to Q/C
interface. Although generated by the basis vectors of three-dimensional
Euclidean space, APS includes a four-dimensional vector space with Minkowski
metric of signatures $\left(  1,3\right)  $ or $\left(  3,1\right)  .$ The
classical eigenspinor and projectors in APS are powerful tools for solving
problems in relativistic dynamics, but their demonstrated close relations to
the quantum Dirac solutions and the implications of these relations for our
understanding of quantum phenomena may be more significant.

A number of classical calculations of components, spin distributions, and
measurements are seen to be fully equivalent to their quantum counterparts. An
eigenspinor analysis of the Stern-Gerlach experiment shows how orthogonal
components of any rotor lead to measurement results given by the eigenvalues
of the measurement operator. The list of Q/C congruences grows when a
classical spin rotation is included as allowed by rotational gauge freedom.
The definition of a classical elementary particle as one whose motion is
described by a single eigenspinor field means that the $g$-factor of such
particles is $2.$ The boosted spin rotation gives de Broglie waves, and a
measurement of their wavelength determines the rotation rate to be at the
Zitterbewegung frequency. The calculated shift in the spin rotation rate in
the presence of a magnetic field gives the magnetic dipole moment, and when
this is combined with the $g$-factor, the magnitude of the spin as $\hbar/2$
is determined. The calculation of the magnetic moment provides new insight
into the magnetic interaction of the spin with an external magnetic field and
shows how the Pauli objection to putting time on the same footing as spatial
coordinates is resolved in APS.

In spite of these many classical associations, there exists an essential core
of quantum behavior that does not seem to have a classical correlate (at least
none has yet been identified). Central to this core is the existence of quanta
themselves and associated Born interpretation\cite{Brum2006} of the particle
current $J$ as a probability current. These are key to the measurement problem
in quantum mechanics and the attendant question of wave-function collapse. The
classical eigenspinor apparently has little to say about these core problems
except to emphasize the central role of amplitudes even in classical
descriptions, and as argued above, the presence of a rapid intrinsic rotation
would require that measurements of elemental quanta see them as point-like
objects. One may also speculate that the generation of APS from
annihilation/creation operators may yield a formulation of measurements that
give quanta in analogy to the measurement of quantized states in a
Stern-Gerlach experiment. In this case, however, the expansion in
\textquotedblleft filters\textquotedblright\ $1=$P$_{3}+$\={P}$_{3}=aa^{\dag
}+a^{\dag}a$ would be imposed not on the left of the spinor, where we saw they
would act as spin-polarization filters [Eq. (\ref{updown})] nor on the right,
where we saw they would select chirality [Eq. (\ref{chiral_split})], but in
some other way yet to be imposed on the formalism.

Some might argue that because our spinor approach emphasizes the role of
amplitudes that can interfere, it is essentially quantum rather than classical
in nature. Indeed, the dividing line at the Q/C interface has become
indistinct in places, but recall that the eigenspinor arises naturally in the
APS treatment of classical dynamics. Furthermore, the analysis presented here
is based on the classical eigenspinor approach rather than on a specific
classical model. We have thus avoided assumptions, for example about the
mechanical structure or charge distribution, that are not required by the
geometry of space. While the lack of a concrete model will be frustrating to
some, it appears necessary in order to ensure a secure mathematical and
experiential footing for our approach. In the geometrical picture that emerges
of spin-1/2 fermions, there are some differences from conventional pictures of
quantum properties. In APS, the spin of any fermion in a pure state is a
vector of length $\hbar/2,$ equal to the magnitude of measured components.
Furthermore, the spin in a pure state has a definite direction with exact
components. This differs from the conventional picture of fermion spin as a
vector of length $\sqrt{3}\hbar/2$ with uncertain components except in the
measured direction. The interpretation suggested by APS is instead that of the
a traditional spin-density operator formulation\cite{Eberly2006,Fano57} or the
description by Levitt\cite{Levitt2001}. The uncertainty relations for the spin
arise here from the measurement process, and the supposed length of $\sqrt
{3}\hbar/2$ arises from the square root of $\mathbf{e}_{1}^{2}+\mathbf{e}%
_{2}^{2}+\mathbf{e}_{3}^{2}$ and the (mis)-identification of the basis vectors
with spin component operators.

We have concentrated here on single-particle systems and have therefore not
discussed the important statistical properties of fermions or the classical
view of entanglement. Much more work is needed. Nevertheless, significant
progress in understanding single-fermion spin in classical terms has been
reported here, and multiple qubit systems have been studied elsewhere with
tensor products of APS,\cite{Cabrera07a} and comparisons of explicit quantum
and classical-eigenspinor solutions with spin have been
undertaken.\cite{CabBay07}

\subsection*{Acknowledgment}

This research was supported by the Natural Science and Engineering Research
Council of Canada.

\end{document}